# High Gradient Silicon Carbide Immersion Lens Ultrafast Electron Sources

Kenneth J. Leedle[1,a)], Uwe Niedermayer[2], Eric Skär[2], Karel Urbanek[3], Yu Miao[1], Payton Broaddus[1], Olav Solgaard[1], Robert L. Byer[3]

[1]Department of Electrical Engineering, Stanford University, Stanford, California 94305, USA
[2]Technical University Darmstadt, Institute for Accelerator Science and Electromagnetic Fields, Schlossgartenstraße 8, 64289 Darmstadt, Germany
[3]Department of Applied Physics, Stanford University, Stanford, California 94305, USA
a)Corresponding author: kleedle@stanford.edu

We present two compact ultrafast electron injector designs with integrated focusing that provide high peak brightness of up to $1.9 \times 10^{12}$ A/m$^2$Sr$^2$ with 10's of electrons per laser pulse using silicon carbide electrodes and silicon nanotip emitters. We demonstrate a few centimeter scale 96 keV immersion lens electron source and a 57 keV immersion lens electron source with a 19 kV/mm average acceleration gradient, nearly double the typical 10 kV/mm used in DC electron sources. The brightness of the electron sources is measured alongside start-to-end simulations including space charge effects. These sources are suitable for dielectric laser accelerator experiments, ultrafast electron diffraction, and other applications where a compact high brightness electron source is required.

## I. Introduction

High brightness ultrafast electron sources are critical for a large variety of applications including microscopy, diffraction, and dielectric laser accelerator (DLA) devices such as free electron lasers. These applications require both a bright electron source and subsequent electron optics to deliver a nanometer size beam to a downstream interaction point. The key electron source figure of merit is the normalized peak brightness $B_{p,n} = J_p/4\pi^2\varepsilon_n^2$ for peak current $J_p$ and normalized rms emittance $\varepsilon_n$ in the beam focus [1]. Research efforts have focused on improving performance from metal tips, coated metal tips, silicon tips, or nanowires in various forms with great success [2-5]. When combined with a meter-scale transmission electron microscope (TEM) column for beam shaping, a laser-triggered tungsten cold field emission source can deliver a $\varepsilon_n=10$ pm-rad beam with a normalized peak brightness of ~$4 \times 10^{13}$ A/m$^2$Sr$^2$ with 0.5 electrons per pulse after filtering out 97% of the electrons [6]. Such a TEM source meets the injection emittance requirements for a scalable dielectric laser accelerator [7,8], but the current is too low for many applications that require multiple electrons per pulse [9,10]. DC sources with flat cathodes can provide 1000's of electrons per pulse with a downstream brightness of ~$2 \times 10^{11}$ A/m$^2$Sr$^2$, but typically have a normalized emittance in the 10 nm-rad range, unsuitable for many applications without emittance filtering [11-13]. Without compression, the beam brightness delivered to a downstream interaction point from DC biased photocathode sources is limited by space-charge effects in the low maximum accelerating field of 10-12 kV/mm [11,12]. Radiofrequency electron guns, on the other hand, can offer accelerating gradients in excess of 100 kV/mm, greatly reducing the influence of space-charge effects on the beam, but generate beam emittances similar to flat DC photocathodes [14]. There have been several demonstrations of very high 50-130 kV/mm DC fields with a flat cathode and hemispherical anode, but these are not commonly used outside of field demonstration experiments, and have not been used in a complete electron source with a nanotip cathode [15,16].

In this paper, we present a compact 96 keV immersion lens [17] electron source that fits within a 25 mm radius footprint with no additional focusing elements and a more flexible high gradient 57 keV immersion lens electron source that includes a separate solenoid lens so that spot size and beam divergence can be optimized for different applications. These sources deliver sub-100 pm-rad normalized emittance beams with peak brightness of up to $1.9 \times 10^{12}$ A/m$^2$Sr$^2$ at a downstream interaction point, and they preserve the full current of the tip source with no emittance filtering. These sources build upon the prototype 30 keV immersion lens source [17], scaling to higher gradients, energy, and beam brightness. Silicon carbide electrodes are used to enable robust turn-key operation at high DC fields, and silicon nanotip cathodes are used to provide an easily integratable nanotip source. Silicon carbide has a high DC breakdown field of 300 kV/mm and is widely used in power electronics for that reason [18]. It is also extremely hard and resistant to heat and many chemicals. Silicon carbide has previously been used in low energy field emission sources and thermionic converters [19,20], but has not been evaluated as a material for electron gun assemblies to the best of our knowledge. The immersion lens electron beams are injected into silicon dual pillar dielectric laser accelerators to characterize their brightness and demonstrate their compatibility with applications that require sub-micron beams with many electrons per pulse [9,10].

## II. Experimental Description

In these experiments, 4H 0.02 ohm-cm nitrogen doped silicon carbide electrodes were laser cut to size and the edges were shaped with diamond abrasive tools and mechanically polished to 0.1 μm finish with diamond lapping film. The maximum field that the silicon carbide electrodes can handle is limited by subsurface defects from the mechanical polishing process. Subsurface defects result in degraded local material properties and are the initiation sites of breakdown. Using chemical-mechanical polishing or similar methods would enable even higher surface breakdown fields. In testing, the SiC electrodes were stable (below 1 fA field emission) in fields of over 100 kV/mm in a simple diode configuration and at peak surface fields of approximately 70 kV/mm for two months of operation in a full immersion lens assembly (failure was caused by mechanical shock to the vacuum system while in operation). The maximum operating gradients in our device were limited by the need for cathode drive laser clearance and the field limits on the silicon cathode from the focusing voltage. The silicon cathode breaks down at surface fields of approximately 40 kV/mm and was the primary source of electron gun failure once most surface defects in the SiC electrodes were eliminated. The high hardness of silicon carbide also facilitated handling and cleaning procedures after polishing compared to metal and silicon electrodes.

Figure 1 shows the 96 keV "Glassbox" immersion lens constructed with a conventional average accelerating gradient of 8.4 kV/mm. The objective for this configuration was to create a ~100 keV DLA electron injector that was as compact as possible, fitting inside of a 25 mm radius "Glassbox". The immersion lens is mounted in a stainless-steel housing suspended inside a 50 mm diameter tube, and faces a 316LN stainless-steel anode at 12 mm distance. A stainless-steel anode was chosen for simplicity with the lower accelerating gradient in this design. The silicon nanotip cathodes are mounted on an adjustable ceramic standoff with a vacuum gap between cathode and the focusing electrode. The immersion lens focus is 13 mm behind the front of the anode, for a total working distance of 25 mm from source to the dual pillar DLA structure. This is slightly longer than the 15 mm working distance for an unfocused 100 keV beam presented in [11], but it provides a beam of 100X smaller emittance.

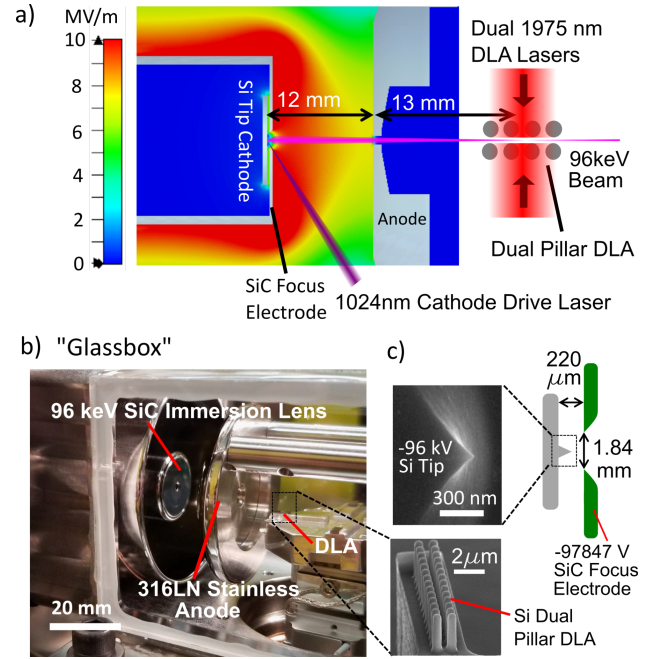

Figure 1. a) Schematic of the "Glassbox" 96keV immersion lens experiment showing electric field magnitude. b) Image of the Glassbox 96keV immersion lens experiment. c) Closeup of the Glassbox immersion lens geometry and scanning micrograph images of the silicon nanotip and dual pillar silicon DLA.

The Glassbox immersion lens used a 360 μm thick silicon carbide focusing electrode with a tapered 1.84 mm diameter aperture and a cathode to focusing electrode gap of 220 μm as shown in Figure 1c. The silicon nanotips used are similar to those used in [17], with a tip height of ~400 nm above the silicon surface and a tip radius of ~20 nm. The cathode chip has an array of nearly identical silicon tips, and the one closest to the center axis is used. Figure 1c shows an image of the silicon nanotip and the silicon dual pillar DLA used for cross-correlation bunch length measurements.

Figure 2 shows the 57 keV "Shoebox" immersion lens source that was designed to provide higher acceleration gradients and a reconfigurable focus size and divergence using the immersion lens together with a separate solenoid. This reconfigurability enabled the immersion lens to be optimized for maximum DLA interaction point brightness. Figure 2a shows that the peak surface field on the silicon carbide electrodes was ~33 kV/mm, which is less than half the breakdown field of these electrodes and resulted in turn-key operation with excellent stability without high voltage conditioning. The 57 keV beam energy was chosen over a 100 keV beam energy primarily because of the reduced electrical and radiation hazards during electron gun failure and is a sufficiently high energy for DLA and ultrafast electron diffraction experiments. The immersion lens in this system is used to set the beam divergence going into the solenoid, and hence the beam's focus size and divergence can be optimized for maximum brightness. The electrical feedthrough for the 57 keV immersion lens was a 40 mm ceramic break, for an overall system size of a shoebox, not including radiation shielding and vacuum pumps. The total working distance for the Shoebox was 213 mm from the cathode to the DLA interaction point, significantly smaller than a typical transmission electron microscope column.

The 57 keV immersion lens was constructed with a 360 μm thick silicon carbide focus electrode with a tapered 1.28 mm diameter aperture and a 2 mm thick silicon carbide anode with a 0.85 mm diameter bore as shown in Figure 2c. The silicon cathode was mounted 110 μm behind the back surface of the SiC focusing electrode. The separation between the SiC focus electrode and the SiC anode was 3.1 mm for these measurements, for a bulk accelerating field of 19 kV/mm. The same geometry silicon nanotips are used as in the Glassbox system, with the closest tip to the center axis used for these measurements.

The silicon nanotips were triggered using five-photon emission with a 1024 nm, 280 fs FWHM laser with a peak intensity of $1\text{-}2\times10^{10}$ W/cm$^2$ at a 30 degree angle to the substrate as illustrated in Figures 1a and 2a. This multi-photon emission process suppresses emission from the cathode substrate as compared to one- or two-photon emission, but results in a larger electron energy spread [21]. The laser repetition rate was 925 kHz for average electron beam currents of up to 7 pA or 47 electrons per pulse as measured with a Faraday cup and electrometer. Tip emission was stable over 100 hours at up to 20 electrons per pulse, but did slowly degrade at above 30 electrons per pulse. Typical emission stability was about 2% rms at 20 electrons per pulse. The operating vacuum was $2\times10^{-9}$ Torr for the Glassbox and $5\times10^{-10}$ Torr for the Shoebox system.

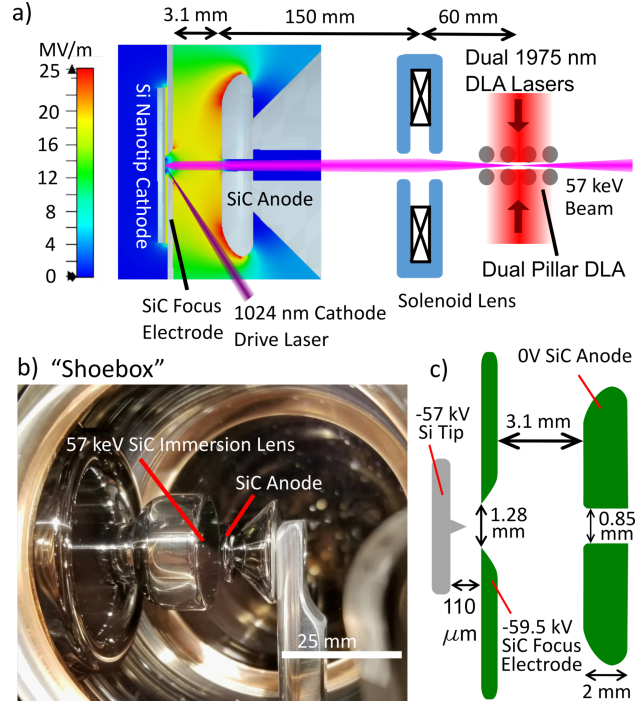

Figure 2. a) Schematic of the "Shoebox" 57 keV experiment showing electric field magnitude in the immersion lens. b) Image of the 57 keV SiC immersion lens. c) Illustration of the 57 keV Immersion lens geometry (not to scale).

### III. Results

Figure 3 shows the raw data collected from the immersion lens sources. First, the unfocused beam profile for <1 electron per laser pulse and 23 electrons per laser pulse were taken for each system using a microchannel plate detector as shown in Figure 3a,b. At higher electron yields, the emission cone from the tip grows, and often forms a ring or crescent moon beam depending on the exact tip parameters. These ring and crescent moon shapes are caused by off-apex electron emission [22]. The focusing voltage and solenoid current were then optimized and the beam emittance was measured by performing knife edge measurements at the focus of the beam on a DLA structure and at a knife edge 26 cm downstream to measure the beam divergence. The knife edge 10% to 90% normalized transmission width was converted to rms width by dividing by 2.56 for an assumed gaussian beam shape. Figure 3c and 3d show example knife edge scans from the 57 kV Shoebox immersion lens. The x and y emittance were within 10% of each other for all the data collected.

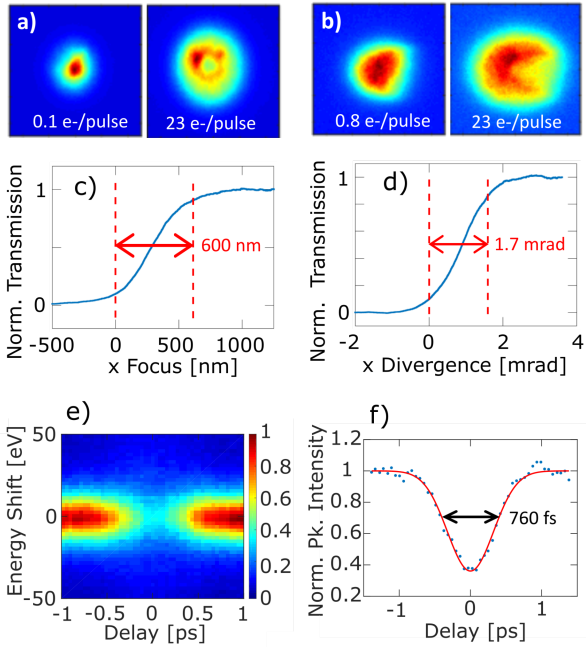

Figure 3. a) Profile of the 96 keV Glassbox beam at low intensity and 23 electrons per pulse with immersion lens turned off (0 $V_{focus}$). b) Profiles of the 57 keV Shoebox beam at low intensity and 23 electrons per pulse with immersion lens and solenoid lens off. c) and d) show example $x_{10-90}$ knife edge focus and divergence measurements from the 57 keV shoebox at the DLA and 26 cm downstream for 1 e-/pulse. Similar measurements were performed for both immersion lens designs. e) Shoebox DLA interaction cross-correlation spectrum vs. delay. f) Peak depletion bunch length measurement at the DLA for the beam shown in c), d), and e).

A dual pillar silicon DLA in the beam focus was used to perform a cross-correlation acceleration experiment to measure the electron bunch length for both systems [18]. Figure 3e shows an example DLA spectrum as a function of delay for the Shoebox system at 1 electron per pulse. Note that the DLA modulates the electrons more than 1 keV so the accelerated electrons are not resolved in this scan. The magnetic spectrometer has a point spread function of 25 eV FWHM. The center peak intensity vs. delay from Figure 3e is shown in Figure 3f and is used to measure the electron bunch length. The gaussian fit FWHM of the center peak depletion is deconvolved with the 310 fs electric field FWHM of the DLA drive lasers to obtain the electron bunch length. Center peak depletion is a more robust bunch length measurement than the width of accelerated electron signal as used in [17] and will usually give longer bunch length measurements. Typical focus intensities for the 1975 nm DLA drive lasers were 6 mJ/cm$^2$ from each side.

Table 1 summarizes the performance of the SiC immersion lenses. The Glassbox 96 keV immersion lens focused the beam onto the DLA at a 25 mm working distance with no additional focusing elements. This source produced a focus size of 410 nm rms, a 0.39 mrad rms divergence angle, a 97 pm-rad normalized emittance, and a bunch duration of 830 fs at low charge. The Glassbox immersion lens produced a maximum peak brightness of $B_{p,n} = J_p/4\pi^2\varepsilon_n^2 = 9.3\times10^{11}$ A/m$^2$Sr$^2$ at $N_e$ = 23 electrons per pulse ($J_p = N_e q_e/t_p$ for electron charge $q_e$ and $t_p$=1230 fs FWHM bunch length) with a 290 pm-rad normalized emittance $\varepsilon_n$. This normalized peak brightness was similar to the $8.6\times10^{11}$ A/m$^2$Sr$^2$ at 28 electrons per pulse obtained with the 30 keV immersion lens prototype [17] which informed the design of the Shoebox system to support higher gradients and better optimization. The maximum average normalized brightness of the Glassbox beam was $B_{avg,n} = J_{avg}/4\pi^2\varepsilon_n^2 = 1.05\times10^6$ A/m$^2$Sr$^2$ in the beam focus for average current $J_{avg}$ = 3.4 pA ($N_e$ = 23 at 925 kHz) and normalized rms emittance $\varepsilon_n$.

Table I: Low charge beam properties

| System | Glassbox | Shoebox |
|---|---|---|
| Beam Energy | 96.0 keV | 57.0 keV |
| Acceleration Gap | 12 mm | 3.1 mm |
| Total Working Distance | 25 mm | 213 mm |
| Focus Voltage Vf | 1847 V | 2500 V |
| Average cathode field [kV/mm] | 1.44 | 5.13 |
| Average Field to anode [kV/mm] | 8.4 | 19.2 |
| Min RMS Focus [µm] (sim) | 0.41 (0.42) | 0.23 (0.21) |
| Min RMS div [µrad] (sim) | 390 (346) | 680 (1000) |
| Min Normalized Emittance [pm-rad] (simulation) | 97 (94) | 77 (102) |
| Min FWHM Bunch Length [fs] (simulation) | 830 (993) | 700 (587) |

The Shoebox 57keV system was designed to be more flexible, with an adjustable gap between the focusing electrode and grounded anode and a separate solenoid lens to focus the nearly collimated electron beam from the immersion lens onto the DLA device. The Shoebox system had optimal performance with a beam divergence of 0.68 mrad rms, delivering 230 nm rms spot sizes, 77 pm-rad normalized emittance, and 700 fs FWHM bunch lengths at low charge. The maximum normalized peak brightness measured was $1.9\times10^{12}$ A/m$^2$Sr$^2$ at 12 electrons per pulse with a 170 pm-rad normalized emittance and a 950 fs FWHM bunch length. This corresponded to a maximum measured average brightness of 1.66x10$^6$ A/m$^2$Sr$^2$ at 1.8 pA (N$_e$ = 12 at 925 kHz). This lower maximum electron current was due to tip deterioration after testing at high electron yields. The Shoebox average brightness is comparable to or better than a CW laser triggered Schottky emitter transmission electron microscope without emittance filtering [1].

Figure 4 shows the beam brightness vs. beam current for the immersion lenses. The beam brightness tends to saturate above a few electrons per pulse due to space charge. The Glassbox system showed a large focus size increase with space charge, but the divergence and bunch length increased more slowly. Focus size, beam divergence, and bunch length all increased more evenly with space charge for the Shoebox. The bulk accelerating gradient did not have a strong effect on the Shoebox beam brightness, since the source brightness was largely limited by the low cathode field in the immersion lens. The performance of the 57 keV immersion lens was similar at 16.5 and 18.4 kV/mm bulk gradients with similar immersion lens focal lengths. The improved brightness of the Shoebox source over the Glassbox source can largely be attributed to its higher cathode field in the immersion lens (5.1 kV/mm vs 1.4 kV/mm, respectively), which was part of the motivation for using a separate solenoid for the DLA interaction point focus.

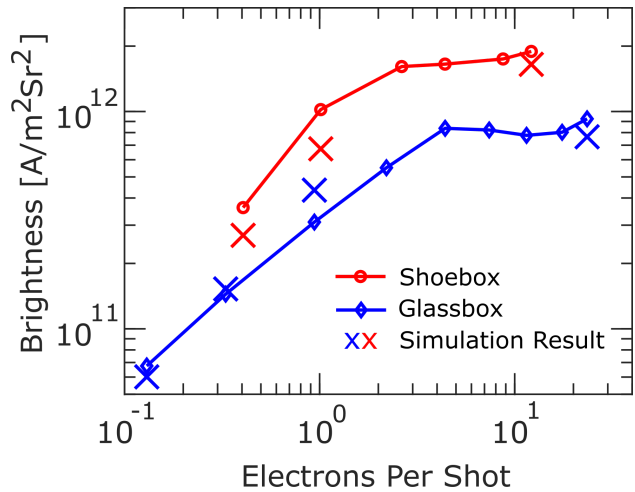

Figure 4. Immersion Lens normalized peak brightness vs. beam current for the 96 keV Glassbox and 57 keV Shoebox systems. Simulation results in the single electron and maximum brightness case are marked with 'x' for each experiment.

The SiC immersion lens devices were modeled using finite element analysis and particle-in-cell tracking simulations in CST Studio to confirm the influence of the bulk accelerating field and cathode extraction field on the source brightness. The computational domain was divided into three parts: the emitter simulation, the immersion lens simulation, and the solenoid simulation in the case of the 57 keV Shoebox system. Figure 5 shows the emitter simulation in a 60 μm (30 μm for the shoebox) long cylindrical region with the boundary potential obtained from the immersion lens simulation. After running through the emitter field the electrons are injected into the immersion lens simulation at z=60μm and z=30 μm respectively. Finally, for the Shoebox there is also magnetostatic simulation performed for the solenoid focus into the DLA.

The tips in both the Glassbox and Shoebox are modeled as a cone of 400 nm height with a base radius of 200 nm and hemispherical apex of 20 nm radius. This results in a field enhancement factor (apex field divided by cathode average field) of 17.5. This enhanced field drops off over 10s of nanometers to the baseline cathode field. The modest 10$^{16}$ cm$^{-3}$ doping of the silicon tips may also reduce the DC field enhancement somewhat due to field penetration [23]. Note that the maximum DC fields on the nanotips are far from the ~GV/m fields required for cold field emission.

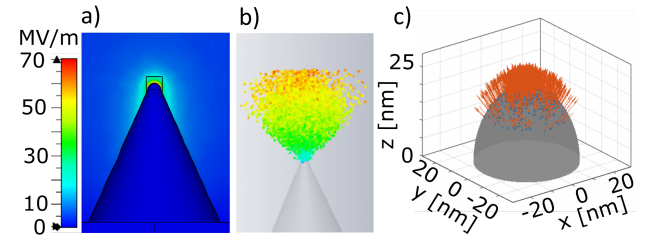

Figure 5. Simulation of silicon tip emission in CST studio. a) DC fields on the tip. b) Emitted electrons (energy color coded) c) Prepared initial distribution (no space charge).

The initial electron distribution is prepared in spherical coordinates with uniform longitude and cosine latitude. The cosine angular distribution is scaled to match the experimental emittance in the absence of space charge. The fitted electron distribution had mean energy of 1.5 eV with a 14.2° rms cosine dependence. The rms energy spread was a fitting parameter to match the bunch length in the absence of space charge and was 0.3 eV for the Glassbox and 1.5 eV for the Shoebox. These are in line with the expected range accounting for multi-photon effects [21] and other noise sources. Unfortunately, the magnetic spectrometer used could not resolve below 25 eV for a direct measurement of the energy spread. Displacing the tip off-center to model alignment error results in strongly asymmetric beam shapes that were not found experimentally, suggesting that the tips used were within 10 microns of the beam axis.

The initial electron temporal profile is taken as Gaussian with 140 fs FWHM, i.e. the length of the laser pulse divided by the square root of the emission photon order. This initial bunch length broadens from 140 fs to ~90% of the final bunch length by the time the electrons reach the focusing electrode due to the starting energy spread and trajectory effects even without space-charge. This broadening can be reduced by increasing the electric field in the cathode region to mitigate chromatic and geometrical broadening when the electrons have low energy. This was also verified experimentally by operating the 57 keV Shoebox immersion lens with zero focusing voltage to increase the cathode field from 5.1 kV/mm at 2500 $V_f$ to 9.3 kV/mm at 0 $V_f$. This decreased the experimental single electron bunch length from 700 fs to 630 fs FWHM, but the emittance and brightness were significantly worse due to aberrations in the solenoid lens.

The simulations were performed at low-charge and at the peak brightness point for each configuration since the full 3D simulations became prohibitively expensive computationally with space-charge dynamics included. The simulated emittance increase had to account for the larger emission cone with increasing current in addition to space charge effects. Increasing the cosine scaling to 18.2° rms matched the experimental peak brightness well for the Glassbox and Shoebox.

**IV Discussion**

The performance of the immersion lenses stays relatively constant over a wide range of device parameters. This is due in large part to the sensitivity of these systems to spherical and chromatic aberrations and coma in the immersion lens itself. Even small shifts of 10 μm of the cathode nanotip from the optic axis cause increases in the achievable focus size. Spherical aberration also strongly increases the achievable focus size downstream of the immersion lens, and hence the nanotip emission spatial profile strongly influences the source emittance. The ~400nm tall silicon nanotips also need to be driven with fourth or fifth order multi-photon excitation to prevent substrate emission. This large photon order contributes to a larger energy spread for the electron source, which is higher than other needle based tip emitters that can be driven at the material work function [24]. Additionally, for a given immersion lens geometry and focal length, changing the bulk dc gradient does not significantly affect the relatively low electric field on the nanotip cathode. A taller nanotip with larger DC field enhancement factor could potentially provide higher performance and less sensitivity to space-charge effects.

**V Conclusion**

The reported SiC immersion lenses enable very compact sources of sub-picosecond electron bunches with excellent beam parameters for dielectric laser accelerator, ultrafast electron diffraction, and other measurements with up to 23 electrons per pulse. We measured normalized emittances smaller than 100 pm-rad and peak brightness larger than $10^{12}$ A/m$^2$/Sr$^2$. The measurements were confirmed by multi-scale simulations in the full 3D fields. For planned scalable DLA experiments with transverse confinement and large energy gain [7,8], these emittances are still about a factor of four too high and further improvement is needed to prevent beam loss. Currently, the sources are limited by the extra energy spread of the silicon emitter in this geometry, as well as spherical and chromatic aberrations. Future designs using different geometry nanotips could take advantage of high extraction field, low energy spread emitters and the high operating gradients enabled by silicon carbide electrodes. Furthermore, silicon and silicon carbide electron source components could be readily integrated into a high-performance accelerator on a chip platform.

**Acknowledgements**

This work was funded by the Gordon and Betty Moore Foundation (No. GBMF4744). We wish to thank T. Hirano and P. Hommelhoff for fruitful discussions.

The authors have no conflicts of interest to disclose.

The data supporting the findings of this manuscript are available from the corresponding author upon reasonable request.